\RequirePackage{fix-cm}
\documentclass[twocolumn]{svjour3}          % twocolumn
\smartqed  % flush right qed marks, e.g. at end of proof
\usepackage{graphicx}
\usepackage{psfrag}

\begin{document}

\title{On the Distributed Construction of a Collision-Free Schedule in Multi-Hop Packet Radio Networks%\thanks{Grants or other notes
%about the article that should go on the front page should be
%placed here. General acknowledgments should be placed at the end of the article.}
}
%\subtitle{Do you have a subtitle?\\ If so, write it here}

\titlerunning{Collision-free schedule in multi-hop packet radio networks}        % if too long for running head

\author{Jaume Barcelo \and
        Boris Bellalta         \and
        Cristina Cano \and
        Azadeh Faridi \and
        Miquel Oliver%etc.
}

%\authorrunning{Short form of author list} % if too long for running head

\institute{
           J. Barcelo \at
           Universidad Carlos III de Madrid \\
           \email{jaume.barcelo@uc3m.es}           %  \\
%             \emph{Present address:} of F. Author  %  if needed
           \and
           B. Bellalta, C.Cano, A. Faridi, M. Oliver \at
           Universitat Pompeu Fabra \\
           \email{\{boris.bellalta; cristina.cano; azadeh.faridi; miquel.oliver\}@upf.edu}           %  \\
}

\date{Received: date / Accepted: date}
% The correct dates will be entered by the editor

\maketitle

\begin{abstract}
This paper introduces a protocol that distributively constructs a collision-free schedule for multi-hop packet radio networks in the presence of hidden terminals.
As a preliminary step, each wireless station computes the schedule length after gathering information about the number of flows in its neighbourhood.
Then, a combination of deterministic and random backoffs are used to reach a collision-free schedule.
A deterministic backoff is used after successful transmissions and a random backoff is used otherwise.
It is explained that the short acknowledgement control packets can easily result in channel time fragmentation and, to avoid this, the use of link layer delayed acknowledgements is advocated and implemented.
The performance results show that a collision-free protocol easily outperforms a collision-prone protocol such as Aloha.
The time that is required for the network to converge to a collision-free schedule is assessed by means of simulation.
% \PACS{PACS code1 \and PACS code2 \and more}
% \subclass{MSC code1 \and MSC code2 \and more}
\end{abstract}

\section{Introduction}
\label{intro}
%Wireless communication is increasingly being used to access the Internet (e.g., \cite{wamser2010tcr}).
Most of the wireless networks in use nowadays rely on some kind of infrastructure.
A clear example is IEEE 802.11 networks in which wireless stations connect to wireless access points and access points are used as gateways to the wired network.
These networks use wireless communications only for the last hop and present some limitations.
To start with, it is difficult to deploy a network in those places where there is no infrastructure available.
It is envisioned that, in the near future, multi-hop packet radio networks should offer networking capabilities in places where network infrastructure is not available.

A packet radio network is defined in \cite{tobagi1987mpa} as a collection of wireless stations that exchange messages via broadcast radio.
Each wireless station (often referred to as wireless node, or simply station or node) consists of a radio and a controller.
The controller is capable of executing routing algorithms and therefore the stations can forward packets that belong to other stations.

Multi-hop packet radio networks present several challenges and interdependencies at all layers of the protocol stack.
Our goal in the present work is to focus on the media access control (MAC) layer and explore the possibility of distributively constructing a collision-free schedule in multi-hop packet radio networks.
A collision is simply the interference caused when there are two active transmitters in the neighborhood of a receiver.
Collisions can be avoided by a careful scheduling of all the transmissions in the network, and the construction of such schedule in a distributed fashion is the object of the present work.

This paper is the extended version of a workshop paper where the main idea was originally outlined \cite{barcelo2011cfo}.
This idea can be summarized as using a deterministic backoff after successful transmissions and a random backoff otherwise.
Each terminal uses random backoffs until it finds a free transmission opportunity where no other terminal is transmitting over the shared channel.
It will then periodically transmit during that same transmission opportunity with a period equal to the schedule length.
If there is a collision, the nodes involved perform random backoffs until they each find another available transmission opportunity.
Once all nodes find their transmission opportunities within the schedule length, there will be no further collisions.

Compared to the original workshop paper, the present work offers the following additional contributions:
\begin{itemize}
\item An extended and updated literature review, which includes the latest advancements in this area.
\item A modification of the protocol to support link layer delayed acknowledgements (ACKs).
\item New simulation results that show that the modified protocol offers a shorter transient state.
\end{itemize}

As mentioned before, the fundamental technique we rely on is the use of a deterministic backoff after successful transmissions and a random backoff otherwise.
This idea has already been used in other problems, as it is described in Sect.~\ref{sec:related_work}, which reviews previous work in this area and offers the necessary background for the remainder of the paper.
The problem of channel time fragmentation that results from the use of link-layer ACKs is introduced in Sect.~\ref{sec:delayed_ack}.
This problem can be alleviated if delayed ACKs are used at the link layer.
Then, in Sect.~\ref{sec:protocol} we cover the protocol that the different stations follow to distributively compute the schedule length and then cooperatively construct a collision-free schedule for wireless transmissions.
Performance results are presented in Sect.~\ref{sec:simulation_results}, where the aggregate throughput and the duration of the transient state is evaluated for different schedule lengths using an example topology.
Further refinements and aspects that are considered out of the scope of this article are discussed in Sect.~\ref{sec:refinements}.
Finally, Sect.~\ref{sec:conclusion} concludes the paper.

\section{Related work}
\label{sec:related_work}

The problems associated to multi-hop packet radio networks (also known as multi-hop radio networks or wireless mesh networks in recent literature) have attracted the attention of the research community for a long time.
As early as 1987, Tobagi presented a survey on the modeling and analysis of multi-hop packet radio networks \cite{tobagi1987mpa}, where he identified the difficulties intrinsic to this kind of networks.
First, because of the broadcast and shared nature of the wireless channel, medium arbitration protocols are needed.
Moreover, the action of each wireless station inevitably affects several surrounding stations, thus incurring interdependencies that complicate the analysis.

To further complicate the issue, wireless propagation introduces an additional degree of randomness and unpredictability to the network.
In order to advance in the analysis of multi-hop packet radio networks, some simplifications are in order.
One of the most common modeling assumptions is the use of a graph representation of the network, where an edge exists between two stations if they are in the transmission range of each other.
As an example, \cite{arikan1984scr} uses a graph representation to show that determining whether a traffic matrix belongs to the capacity region of a packet radio network is NP hard.
We will use the graph topology representation and several other assumptions in the present paper.
%We explicitly leave out of the scope of the present paper the queueing dynamics and multi-packet transmission and reception \cite{bellalta2012rqp}.

Our goal is that the different stations of the network settle down in a satisfactory transmission schedule that efficiently uses the radio resources.
In \cite{ramanathan1993sam} the problem is treated from a theoretical perspective in which the network topology is known and the schedule can be constructed in a centralized way.
An interesting alternative is considered in \cite{yi2010msl} were contention MAC protocols are proposed in which the stations learn from their neighbourhood.
These two works rely on a slotted channel access, which means that neighbouring stations need to have closely synchronized clocks.
The problem of gradient clock synchronization is treated in \cite{sommer2009gcs}.

A possible solution to avert the problems of network synchronization and schedule construction is to use random media access control.
If all the stations use the Aloha protocol, the contention parameter can be optimized to maximize proportional fairness as demonstrated in \cite{kar2004apf}.
The contention parameter regulates how aggressively an Aloha station contends for the medium.
If the contention parameter is too aggressive, a large fraction of channel time is wasted in the form of collisions.
If it is not aggressive enough, a large fraction of the channel time remains idle.

Another alternative, which is the one considered in the present work, is the use of learning MAC protocols to distributively construct a collision-free schedule.
This approach has already been exploited in the area of wireless local area networks \cite{barcelo2008lba}.
Specifically, \cite{barcelo2008lba} suggests the use of a deterministic backoff after successful transmissions to prevent collisions.
The advantage of \cite{barcelo2008lba} compared to previous solutions to prevent collisions (e.g.,\cite{choi2005eei,tian2008ipc}) is that it requires minimum changes to the protocol and it is backward compatible with existing implementations \cite{barcelo2010fcc}.

Later works have explored the validity of the idea for different kinds of traffic \cite{bellalta2009vtc}, modelled the network performance metrics \cite{barcelo2009cpa}, and studied the possibility of traffic differentiation \cite{barcelo2009tpc}.
A comprehensive simulation study, together with a model of the learning process is presented in \cite{he2009sbr}.
This last paper also includes performance measures in non-ideal conditions, such as the presence of legacy stations or channel errors.
More results regarding the learning process and fairness with legacy stations can be found in \cite{barcelo2010fcc}.
An evaluation of a collision-free protocol and its interaction with the autorate fallback mechanism (ARF) in realistic multiple-input-multiple-output (MIMO) channels is presented in \cite{martorell2012pec}.

A general discussion of this class of protocols, together with performance comparisons and protocol refinements, is offered in \cite{fang2010dlm}.
This last paper also introduces the concepts of stickiness and variable schedule length to accommodate a larger number of contenders in slotted networks.
In the present work, we will reuse both concepts in the context of multi-hop packet radio networks.

Collision-free MAC protocols for multi-hop packet radio networks have already been considered in \cite{hui2011epp}. 
However, it was under the assumption of slotted channel time.
In the present paper we study networks in which slot synchronization is not available.
%There has also been interest on collision-free protocols in the area of sensor networks \cite{khan2011cfm}.

The general problem of constructing a collision-free schedule belongs to the family of decentralized constraint satisfaction problems \cite{duffy2011dcs}.
Other problems of this family include channel selection in WLANs and inter-session network coding.
The same principle can be used for channel selection in  cognitive networks as in \cite{khan2011sod}.

From a more practical point of view, many of the multi-hop packet radio networks deployed so far (e.g., \cite{chambers2002grr} and \cite{oliver2010wca}) use IEEE 802.11 equipment that relies on carrier sense multiple access with collision avoidance (CSMA/CA)  for media access control.
However, it has been shown that CSMA/CA performs poorly in multi-hop packet radio networks because potential interferers are often beyond the carrier sense range \cite{gurewitz2009mmo}.

\begin{figure}[]
\centering
  \includegraphics[height=0.25in]{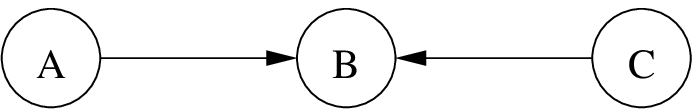}
(a)

\vspace{0.3in}
  \includegraphics[height=0.25in]{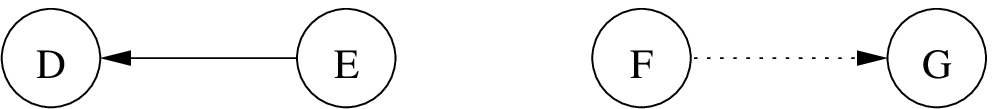}
(b)

\vspace{0.3in}
  \includegraphics[height=0.25in]{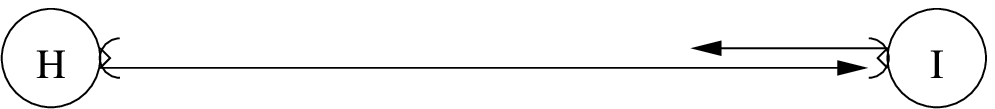}
(c)
\caption{Problems arising when using CSMA in multi-hop packet radio networks and long-distance wireless networks. (a) the hidden terminal effect. (b) the exposed terminal effect. (c) the distant terminal effect.}
\label{fig:problems}
\end{figure}

In Fig. \ref{fig:problems} we describe three of the problems that are associated with the use of CSMA/CA in multi-hop packet radio networks.
The stations are labelled circles and transmissions are represented as arrows.

The first problem is the \emph{hidden terminal problem} (Fig. \ref{fig:problems}(a)) which occurs when two stations that cannot carrier-sense each other simultaneously transmit.
The intended recipient of the transmissions is a third station (labelled $B$ in the figure) that cannot recover either one of the two transmissions, since they overlap in time.

A complementary problem is the \emph{exposed terminal problem} which arises when two stations avoid to transmit simultaneously even though they do not interfere with each other.
This is illustrated in Fig. \ref{fig:problems}(b), where station $F$ wants to transmit to $G$ and it is waiting for $E$ to finish.
In this particular scenario, $D$ and $F$ are not in each other's transmission range, and the same is true for $E$ and $G$.
Therefore, the fact that $F$ is deferring its transmission implies an inefficient use of the radio channel resources.

The last problem is the \emph{distant terminal problem} which is depicted in Fig.~\ref{fig:problems}(c).
The two stations $H$ and $I$ are separated by a very long distance.
At a given time $H$ senses the channel idle and starts a transmission.
While the radio signal is traversing the long distance from $H$ to $I$, station $I$ still senses the channel idle and starts a transmission to $H$.
When $H$'s signal finally arrives to $I$, station $I$ is transmitting and therefore $H$'s transmission is lost due to collision.

As CSMA suffers from these three inefficiencies, the protocol that we suggest in Sec. \ref{sec:protocol} does not make use of carrier sense information to build the collision-free schedule for multi-hop packet radio networks.

The interest on multi-hop packet radio networks has resulted in a standardization effort described in the IEEE 802.11s standard amendment \cite{hiertz2010wms}.
Regarding the MAC layer, IEEE 802.11s offers two alternatives: the compulsory Enhanced Distributed Channel Access (EDCA) and the optional Mesh Coordinated Channel Access (MCCA).

EDCA is a CSMA mechanism that suffers from the aforementioned three inefficiencies.
An assessment of the performance of an IEEE 802.11s network that supports voice-over-IP (VoIP) flows is presented in \cite{andreev2010ssv}.
%An implementation of IEEE 802.11s that is available in NS-3 and it is used in \cite{andreev2010ssv} to assess the performance of a packet radio network that supports VoIP flows.

To prevent the hidden terminal problem, MCCA uses channel reservation.
However, reservation is not the ultimate solution to interference.
Firstly, the reservation needs to be done on the same channel that is in risk of collisions due to the hidden terminals.
Therefore, it is likely that the reservation suffers a collision and consequently the data packets also collide.
Secondly, in the case of successful reservation, interference can still occur because of layer 2 ACKs \cite{krasilov2011iem}.
We will devote some discussion to ACKs in the next section and suggest the use of delayed ACKs to alleviate this problem.
Thirdly, another source of interference is the presence of traffic flows beyond the two-hop neighbourhood \cite{cicconetti2008sdr}.
In that paper it is suggested to change the MCCA schedule when interference is detected.
Our approach also combats this kind of interference, since any unsuccessful transmission triggers a schedule change until collision-free operation is reached.

Compared to previous work (see Table \ref{tab:related_work}), our proposal is the first that simultaneously satisfies the following three requirements:
\begin{itemize}
\item It achieves collision-free operation.
\item It operates in multi-hop packet radio networks.
\item It does not require a slotted channel.
\end{itemize}

This paper presents the basic core principles of collision-free operation in multi-hop packet radio networks.
We describe a bare-bones protocol to illustrate the possibility of collision-free operation in networks in which the channel time is not slotted.
There are many possible refinements to be made on top of this minimalist protocol and we mention some of the most promising lines of research in Sect. \ref{sec:refinements}.

\begin{table}[]
% increase table row spacing, adjust to taste
%\renewcommand{\arraystretch}{2.0}
%
\caption{This table presents a comparison with previous work. Even though \cite{hiertz2010wms} uses slots, we mark it as non-slotted as the slots in IEEE 802.11s are of local significance only. }
\label{tab:related_work}
\centering
\begin{tabular}{|c| |c|c|c|}
\hline
  paper & collision-free & multi-hop & non-slotted  \\
\hline
  \cite{barcelo2008lba},\cite{fang2010dlm} & yes & no & no  \\
\hline
  \cite{hui2011epp} & yes & yes & no \\
\hline
  \cite{hiertz2010wms} & no & yes & yes \\
\hline
  present paper & yes & yes & yes \\
\hline
\end{tabular}
\end{table}

\section{Link layer delayed ACK in multi-hop packet radio networks}
\label{sec:delayed_ack}

Given the uncertainty of wireless communications, it is common in wireless networks for each link layer unicast transmission to be followed by an ACK.
The problem is that, in the presence of hidden terminals, this short ACK control frame can interfere with long data packets transmitted by other terminals hidden to it, thus causing a collision and a considerable waste of channel time.

\begin{figure}
\centering
  \includegraphics[width=2.3in]{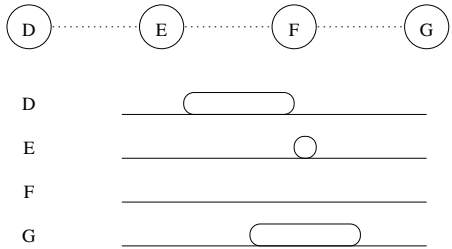}
\caption{A short ACK packet sent by $E$ interferes with the packet that is being received at $F$.}
\label{fig:killer_ack}
\end{figure}

To illustrate this problem we use a topology with four stations in a row as in Fig. \ref{fig:killer_ack}, where stations that are within hearing range of each other are connected with a dashed line.
Station $D$ starts a transmission intended for station $E$.
Some time later, $G$ starts a transmission intended for station $F$.
As soon as $D$'s transmission finishes, $E$ replies with a short ACK.
This short control frame interferes with $G$'s transmission which is not correctly received by station $F$ and therefore, it is not acknowledged.

The fact that each unicast data frame needs to be acknowledged, duplicates the number of transmissions in the packet radio network.
Furthermore, a packet needs to be retransmitted if either the data or acknowledgement frames are lost.
This means that the mere existence of ACKs increases the chances of collisions and retransmissions and, as a consequence, reduces the network performance.

This problem can be partially alleviated by using delayed ACK techniques at the link layer.
Delayed ACK is commonly used in transport control protocols (TCP) to reduce overhead.
The idea is that the ACK is delayed and piggybacked to a data packet.

We suggest the use of delayed ACK at the link layer to prevent \emph{channel time fragmentation}.
By channel time fragmentation we mean that it is difficult or impossible to find the contiguous amount of channel time required for a successful transmission.
Fragmentation is a common issue in computer file systems and is mentioned in the context of multi-hop packet radio networks in \cite{cicconetti2008sdr}.

\begin{figure*}
\centering
  \includegraphics[height=1.5in]{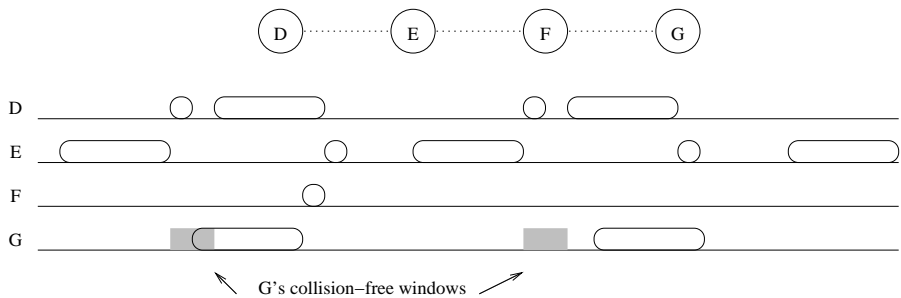}
\caption{$G$'s collision-free windows are those intervals at which $G$ can start a transmission to $F$ without being interfered by $E$'s transmissions or ACKs. In the figure, collision-free windows are represented as a shaded area.}
\label{fig:channel_fragmentation}
\end{figure*}

To illustrate the problem of channel fragmentation we propose the following example.
In Fig. \ref{fig:channel_fragmentation}, stations $D$ and $E$ are sending (and acknowledging) packets to each other.
For the purpose of this example, we assume fixed transmission duration.
ACKs are transmitted right after the data frame is successfully received.
The figure also shows $G$'s \emph{collision-free windows}, which are the time interval at which $G$ can start a transmission to $F$ that will not be interfered by $E$'s transmissions or ACKs.
The collision-free windows are represented as shaded areas in the figure.
As an example, the figure shows two transmissions by $G$.
The transmission that starts in one of the collision-free windows is successful and is acknowledged.
The second transmission by $G$, which starts outside of the collision-free windows, collides with an ACK transmitted by $E$ and is lost.
Notice that $G$'s collision-free windows represent only a small fraction of the channel time.

\begin{figure*}
\centering
  \includegraphics[height=1.5in]{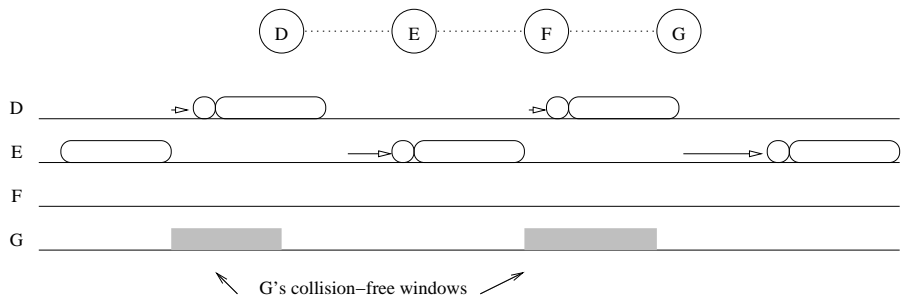}
\caption{$G$'s collision-free windows grow when ACKs are delayed, avoiding channel fragmentation.}
\label{fig:delayed_ack}
\end{figure*}

Compare the situation with Fig. \ref{fig:delayed_ack} where delayed ACKs are used.
The stations do not transmit ACK frames immediately.
Instead, each station waits until it transmits a data frame and prepends the ACK to the data frame.
As a result, $G$'s collision-free windows more than double, which is a very positive outcome.

It can also be the case that a station has several ACKs to transmit when it accesses the channel.
If this is the case, the station transmits all the ACKs before transmitting its own data.
Transmitting the ACKs and the data consecutively has some advantages in terms of overhead, since it avoids the need for dedicated inter-frame spaces, training sequences and headers for the ACK control packets.

In this work we study a saturation scenario, in which all the stations have always a data packet ready to be transmitted.
For completeness, it is also necessary to mention the case in which a station has one or more ACKs to transmit and no data packets when it completes its backoff.
That station is expected to transmit the ACKs, even though it has no data packet.

In the remainder of the paper, the use of link layer delayed ACK is assumed.
%Since ACK information is much smaller than the amount of information included in a regular data packet, we neglect the increase of the length of the data packets that results from the fact that one or more ACKs have been piggybacked.

\section{Constructing a collision-free schedule in multi-hop packet radio networks}
\label{sec:protocol}

Collision-free schedule in multi-hop packet radio networks is advantageous in the sense that no channel time is wasted in collisions and the behaviour of the network is more predictable.
A possible alternative to construct such collision-free schedule is to gather all the information of the network in a centralized point, construct a collision-free schedule, and then disseminate this schedule to the network stations.
%A caveat is that, in order to gather and disseminate information, the network needs to be operative in the first place.

Another alternative is to distributively construct the collision-free schedule in such a way that each station takes its own decisions using information gathered from its immediate neighbourhood.
This is the approach that we use in this paper, and the protocol that we propose has two clearly differentiated phases: the first one is the computation of the schedule length and the second one is the construction of the collision-free schedule.

We will be working with saturation and bounded transmission duration assumptions in the following.
Saturation means that, for each one-hop data flow, the transmitter has always a packet ready to be transmitted to the receiver.
This assumption can be justified by the fact that it represents the maximum load that can be placed on the network.
If the network is not saturated, it means that there is more bandwidth available than the one that is actually required, and therefore the network operation is satisfactory.
As soon as queues build up, the saturation assumption holds.

Regarding the bounded transmission duration, each station is granted a \emph{transmission opportunity} each time it accesses the channel.
A station is not allowed to occupy the channel for a time longer than the transmission opportunity.
Note that a station does not have to necessarily use all of its transmission opportunity as it is just an upper bound on the time that it can access the channel.

%The existence of a maximum transmission length is also a common assumption, since MAC protocols usually do not allow participating stations to take the channel indefinetely.
%We will often refer to the maximum transmission length as transmission length, for brevity.

The concept of transmission opportunity is not new and it is used in IEEE 802.11 EDCA.
The existence of a transmission opportunity is natural in MAC protocols to prevent that a single station captures the channel for an extremely long time.
%The abidance of the stations with the protocol is enforced in the certification process in the case of IEEE 802.11.
For convenience, in this paper we will normalize the duration of a transmission opportunity to one.
It should be clarified that a station has to transmit both the ACKs for the previously received packets and the data within its transmission opportunity.

\subsection{The schedule length}

The goal of the protocol is to construct a collision-free schedule by trial and error.
Using that schedule, each station knows exactly which is the right time to transmit without causing or receiving interference.
Once the collision-free schedule has been constructed, it repeats in time in a periodic fashion.
Each of the participating stations is given a chance to transmit at least once within the duration of the schedule.

The schedule length needs to be long enough to accommodate all incoming and outgoing flows in the neighbourhood of a station.
In fact, we require the schedule length to be strictly larger than the total number of flows in the neighbourhood, multiplied by the length of the transmission opportunity.
This ensures that, given a randomly constructed schedule, there is a non-zero probability that this schedule is collision-free.

Note that the schedule length, as described above, does not need to be the same for all stations.
Those stations that have many flows in their neighbourhood may require a longer schedule length than stations having only a few flows in their neighbourhood.
Still, for the protocol to behave as intended once a collision-free schedule is constructed, it is necessary that the behaviour of the whole network is periodic with a period equal to the longest of the different schedule lengths.
In the following we describe an ingenious trick that has already been used in other papers (e.g, \cite{fang2010dlm}, Sect. 5.3).

The goal here is to choose the schedule lengths in such a way that for any two different schedule lengths,the longer one is an integer multiple of the shorter one.
%Still, we will require that for any two different schedule lengths in the network, the longer schedule length is an integer multiple of the shorter one.
To this extent, we will compute each schedule as an integer power of two multiplied by the duration of the transmission opportunity plus an arbitrarily small finite value ($\epsilon$).
This small value is needed to guarantee that the schedule length computed by each node is strictly larger than the time consumed by the flows in its neighbourhood.

For convenience, we normalize the transmission duration to one, and then we write
\begin{equation}
T_{\sigma_i} = 2^{n_i} (1+\epsilon),
\label{eq:distributed}
\end{equation}
where $T_{\sigma_i}$ is the schedule length for station $i$, $\epsilon$ is an arbitrarily small positive value, and $n_i$ is an integer that we discuss next.
We need $2^{n_i}$ to be larger than the total number of flows in $i$'s neighbourhood, and thus we choose $n_i$ as
\begin{equation}
n_i = \left \lceil \log_2 \left( \sum_{k \in K_i} |I_k| + |O_k| \right) \right \rceil,
\label{eq:integer}
\end{equation}
where $\lceil \cdot \rceil$ is the ceiling operator and $K_i$ is the set of neighbours of station $i$. $I_k$ and $O_k$ are the number of incoming and outgoing flows of station $k$, respectively.

If we compute $n_i$ as in (\ref{eq:integer}), then $2^{n_i}$ is the smallest power of two not smaller than the number of flows in the neighborhood of station $i$.

We use powers of two and logarithms of two in our approach.
Any integer larger than two can be used instead of two.
However, using two gives the finest granularity \cite{fang2010dlm}.

Note that (\ref{eq:distributed}) is constructed in such a way that if $T_{\sigma_i} < T_{\sigma_j}$ for some pair of stations $i,j$, then $T_{\sigma_j}$ is an integer multiple of $T_{\sigma_i}$, which means that the behaviour of station $i$ is also going to be periodic with period $T_{\sigma_j}$.
Since all the schedules evenly divide each other (just as in \cite{fang2010dlm}, Sect. 5.3), the behaviour of the network is periodic as a whole.
The global network period is simply 

\begin{equation}
T_{\sigma}=\max_{i} T_{\sigma_i}.
\end{equation}

The computation of $T_{\sigma_i}$ is very conservative in the sense that it does not take into account the fact that some flows in the same neighbourhood may overlap in time without interference.
As an example, looking at Fig. \ref{fig:problems}(b), transmissions from $E$ to $D$ and from $F$ to $G$ can overlap on time.
If we take station $E$ and compute the number of flows in its neighbourhood, we observe an incoming flow in $D$ and an outgoing flow in $F$.
These count as two different flows even though they can overlap in time.
Furthermore, a flow between two neighbours will be counted twice.

These factors can result in the computation of a schedule length that is longer than what it is strictly necessary.
Even though a longer schedule translates into a lower efficiency in the steady state, it also a shortens the time required to convergence to collision-free operation.

The fact that different stations use different schedule lengths introduces some difficulties when link layer delayed ACKs are used.
A station that transmits to a neighbour with a larger schedule length may not receive its ACK in time, even if no collisions occur.

If two stations use the same schedule length, they will transmit alternatively one after the other and the ACKs will always arrive before the deterministic backoff expires.
This is not necessarily true when the two stations use a different schedule length, since one station may transmit more often than the other.
When different schedule lengths are used, it is necessary to introduce stickiness to the backoff protocol.
Stickiness is a property of learning protocols (e.g., \cite{fang2010dlm}) which is discussed in subsection \ref{subsec:stickiness}.

The suggested approach to compute the schedule length still requires that the stations exchange a limited amount of information with their immediate neighbours before the schedule is computed.
Indeed, it requires exactly the same information as is required for the computation of the contention parameter that maximizes proportional fairness when the Aloha protocol is used \cite{kar2004apf}.
It should be possible to convey this information in short beacons or \emph{hello} messages before the schedule is computed.
This information can be exchanged when the network starts to operate, and consists of very short signaling packets.

\subsection{The backoff protocol}

After each station has computed its schedule length, it will use a very simple protocol to contend for the medium.
This protocol does not use carrier sense information. 
It relies exclusively on the outcome of the last transmission attempt, 
either success or failure, to schedule the next transmission attempt.

It will defer the first transmission for a backoff time that is drawn from an exponential distribution with a parameter equal to the schedule length.
When the exponential backoff expires, the station is allowed to transmit.
The station is allowed to occupy the channel for the length of a transmission opportunity.
Even if the station uses only a fraction of the transmission opportunity, it has to wait until the end of the transmission opportunity before proceeding with the next step of the protocol.

After the transmission opportunity ends, the station will wait for a deterministic backoff equal to the schedule length minus the duration of the transmission opportunity.
When this deterministic backoff expires, the station will be granted a new transmission opportunity if the last packet has been already acknowledged.
Otherwise, it will wait for an additional exponential backoff before starting the transmission opportunity.

\begin{figure}
\centering
  \includegraphics[width=\linewidth]{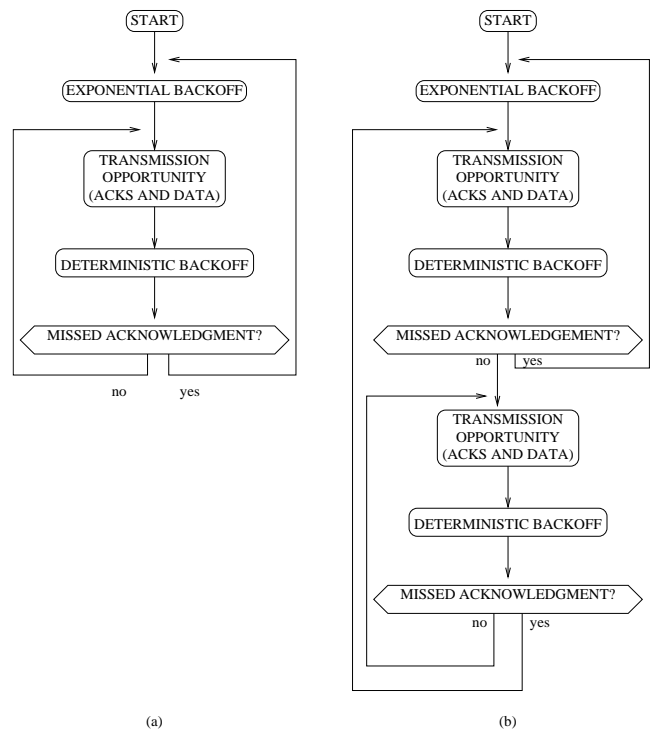}
\caption{Flow chart description of the backoff protocol executed by each station. Regular version (a) and the sticky version with a degree of stickiness equal to two (b).}
\label{fig:flow_chart}
\end{figure}

It is important to highlight that the time that elapses from the start of a transmission opportunity to the next transmission opportunity is completely deterministic and exactly equal to the schedule length $T_{\sigma_i}$ of the station, if both the data transmission and the ACK are successful.
The operation of the protocol is summarized in the flow chart in Fig.~\ref{fig:flow_chart}(a).

In each cycle, there is a transmission opportunity that each station can use to send ACKs and data.
The station must go through the transmission opportunity step in all cases (even if it has nothing to transmit) and the duration of this step is deterministic.
The next step is the deterministic backoff with a duration equal to the schedule length minus the transmission opportunity.
Therefore, the step of the transmission opportunity plus the step of the deterministic backoff has a deterministic duration which is equal to the schedule length.
When there are no transmission errors, no collisions and no missed acks, the duration of the whole cycle is deterministic.
Only when there are collisions or errors, the additional step that extends the backoff for a random amount of time is included.

The random behaviour introduces a change in the schedule that may resolve the conflicts. The underlying philosophy is that when the schedule is not collision-free, there is at least one station that behaves randomly to force a schedule change.
When a collision-free schedule is reached, the behaviour of all the nodes becomes deterministic to stay in that collision-free schedule.

\begin{figure*}
\centering
  \includegraphics[height=1.5in]{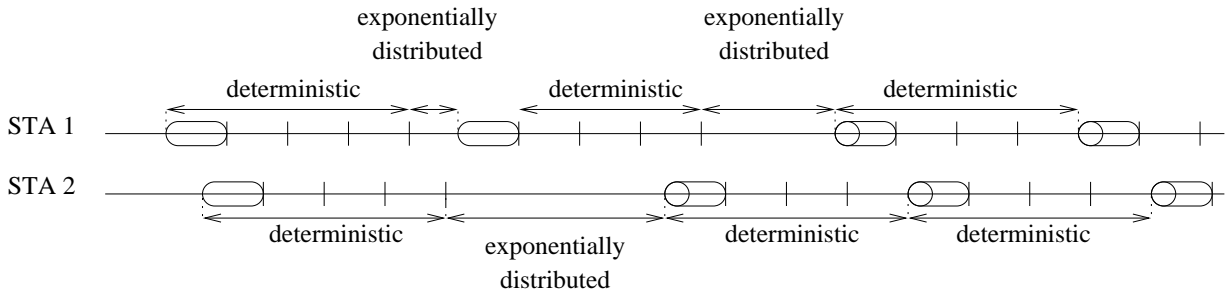}
\caption{An example of the operation of the protocol with two contending stations.}
\label{fig:L-aloha}
\end{figure*}

The operation of the protocol is exemplified in Fig.~\ref{fig:L-aloha}.
STA1 and STA2 are contending for the medium to transmit packets to each other  and collide in their first transmission attempt.

They are using a schedule length equal to four time units, and the transmission opportunity duration is one time unit.
After they complete the transmission opportunity, they back off for a deterministic time equal to three units of time (the schedule length minus the transmission opportunity duration).
Neither one of the two stations has received a positive ACK by the time they finish their deterministic backoff.
Consequently, they back off for an additional random time that is exponentially distributed.

STA1 succeeds in its next transmission attempt.
After the transmission opportunity, STA1 deterministically backs off for three units of time.
When the deterministic backoff is completed, STA1 has not received any ACK packet, and therefore it has to extend its backoff period for a random amount of time.
During the random backoff of STA1, STA2 finishes its own random backoff and successfully transmits.
This last transmission includes an ACK for STA1, which is denoted in the figure by a small circle at the beginning of the transmission.
Then STA1 successfully transmits, including an ACK for STA2.
From this point on, the behaviour of the system is completely deterministic and periodic.
The period is equal to the schedule length, and the systems operates in a collision-free fashion.

%When the collision-free schedule is reached, the stations use a deterministic backoff even though they have to wait for the delayed acknowledgement.
%The delayed acknowledgement arrives always before the deterministic backoff expires.

In our example, when the collision-free operation is reached, the ACK for a given packet arrives always before the deterministic backoff of the packet's transmitter expires.
Therefore, the packet's transmitter is allowed a new transmission opportunity as soon as its deterministic backoff expires, and the system keeps working in a deterministic fashion.

\subsection{Stickiness}
\label{subsec:stickiness}

The example provided in Fig. \ref{fig:L-aloha} illustrates a simple case in which the transmitter and the receiver share the same schedule length.
In complex topologies, it can also be the case that different stations use different schedule lengths.
If that is the case, a station may miss an ACK after the deterministic backoff has elapsed because of the fact that the receiver is using a longer schedule length.
A missed ACK will be followed by a random exponential backoff that moves the station back to its random behaviour.
To prevent this situation, the protocol needs to be modified to accept that some of the ACKs may be delayed.
This can be accomplished by using a sticky variant of the protocol.
Stickiness has already been used in \cite{fang2010dlm} and \cite{barcelo2011tcf} in the context of slotted networks, with the aim of reducing the duration of the transient state.
A sticky protocol with a degree of stickiness equal to two will continue to operate deterministically even if one out of every two ACKs is missed.
The flowchart of the sticky protocol (with a stickiness degree equal to two) is depicted in Fig.~\ref{fig:flow_chart}(b).
Depending on the topology of the network, a larger degree of stickiness might be needed.

If stickiness is applied, it would be applied to all the transmissions, regardless of the destination node.
The use of stickiness has some implications on the overall behaviour of the MAC protocol.
As an example, stations would have multiple transmitted and not acknowledged packets.
The acknowledgement timeouts should be adapted accordingly and the additional delays would affect the time required to converge to collision-free operation.
For example, if a stickiness degree equal to two is used, the time-out has to be set to two times the schedule length.
This change will necessarily increase the reaction time to detect and correct schedule configurations in which collisions occur.

\subsection{Convergence to collision-free operation}
If the interference graph of the network topology is strongly connected, it is guaranteed that the collision-free schedule will be reached in finite time \cite{barcelo2011cfo}.
In a schedule that is not collision-free,  there is at least one station that is behaving randomly because it does not receive ACKs for its packets.
This station will change the position of its transmission opportunity within the schedule until a collision-free schedule is reached.

If that is not possible, the station will keep using a random backoff and, sooner or later, as a result of either delay or interference, some other station will miss one or more ACKs and move back to the random mode of operation and use random backoff.
Since this can result in a cascading effect, a station that behaves randomly can trigger a schedule change that affects all the stations in the network.
Therefore, the only stable state is the collision-free operation.

The interference graph is always strongly connected  when data flows are bidirectional, otherwise additional measures should be taken.
A possibility is to require stations to keep track of the data flows addressed to them, and use a random backoff if they do not receive such flows for the duration of several schedule lengths.
The goal is to ensure that from any state of the network, it is possible to find a sequence of transmissions and collisions that drive the network to a collision-free operation.

For a given topology, to show that the system will eventually converge to collision-free operation we require that, for any possible state of the system $c$, there exists a chain of events of finite duration $t_c$ which can bring the system to collision-free operation with non-zero probability $p_c$.
Let $t_c^*$ denote the largest $t_c$ and $p_c^*$ the minimum $p_c$ over all possible states $c$.
Then, for any given initial state of our network, the probability of convergence is at least $p_c^*$  after a time interval $t_c^*$.
After $N$ time intervals, the probability of the network not having converged is not larger than $(1-p_c^*)^N$ and, as $N$ goes to infinity, the probability that the network has not converged goes to zero.
Consequently, the network converges with probability one.

The proof for the general problem of decentralized constrained satisfaction that directly applies to slotted scenarios is presented in \cite{duffy2011dcs}.

\subsection{Stations transmitting multiple flows}
So far we have assumed that each station is transmitting data to another station.
It is common in multi-hop packet radio networks that a station has data to transmit to several neighbours.
In that case, an instance of the backoff protocol should be executed for each of the destinations.
This is similar to the different backoff instances used by different traffic classes in EDCA.
As in EDCA, having multiple backoff instances may result in internal collisions that can be easily resolved.

We have already considered the possibility that a station has different outgoing flows in the computation of the schedule length.
The parameter $O_i$ in (\ref{eq:integer}) is precisely the number of outgoing flows of station $i$.

Note that the protocol proposed in the present paper is different from the one in the original workshop paper \cite{barcelo2011cfo}.
The motivation of the changes is to take into  account the existence of ACKs.
The results presented in the next section show that the protocol suggested here converges faster to a collision free operation than the original one.
The behaviour of the present protocol in the steady state is exactly the same as the one in \cite{barcelo2011cfo} and for this reason we can reuse the steady state performance metrics of that paper.

\section{Performance and simulation results}
\label{sec:simulation_results}

Since the transmission opportunity is normalized to one, the fraction of channel time used for transmission by a station $i$ in the steady state is
\begin{equation}
\theta_i= \frac{O_i}{T_{\sigma_i}},
\label{eq:throughput}
\end{equation}
where $O_i$ is the number of outgoing flows in station $i$ and $T_{\sigma_i}$ is the schedule length of that station.

\begin{figure}[]
\psfrag{d_1}[cc][cc]{$s_1$}
\psfrag{d_2}[cc][cc]{$s_2$}
\psfrag{d_3}[cc][cc]{$s_3$}
\psfrag{f_1}[cc][cc]{$f_1$}
\psfrag{f_2}[cc][cc]{$f_2$}
\psfrag{f_3}[cc][cc]{$f_3$}
\psfrag{l_1}[cc][cc]{$l_1$}
\psfrag{l_2}[cc][cc]{$l_2$}
\centering
\includegraphics[width=2.9in]{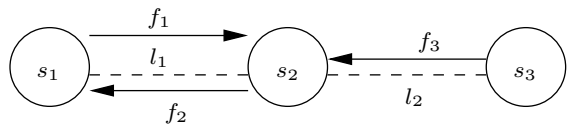}
\caption{Topology of a simple packet radio network.}
\label{fig:basic}
\end{figure}

For illustration purposes, we will use the topology in Fig.~\ref{fig:basic} with three stations, two links and three data flows.
Each station sees a total of three flows in its neighborhood.
Using (\ref{eq:distributed}) and (\ref{eq:integer}) we obtain that the schedule length for all the stations in this particular example is \mbox{$4(1+\epsilon)$}.
Then we can use (\ref{eq:throughput}) to compute the fraction of channel time devoted to successful transmissions by each of the stations and other network performance metrics (fairness, aggregate throughput, and proportional fairness).

In Table \ref{tab:comparison} we compare the performance of the protocol described in the previous section, that we call self-configuring learning Aloha (scl-Aloha) to the Aloha protocol configured with optimal contention parameters to maximize proportional fairness \cite{kar2004apf}.

Table \ref{tab:comparison} shows the values of $\theta_1$, $\theta_2$ and $\theta_3$ for both Aloha and scl-Aloha.
The details on the computation of the values for Aloha can be found in the appendix in \cite{barcelo2011cfo}.
The last three columns represent three different performance metrics:
Jain's fairness index \cite{jain1991acs}
\begin{equation}
JF=\frac{ \left( \theta_1 + \theta_2 + \theta_3 \right)^2}{3 \left( \theta_1^2 + \theta_2^2 + \theta_3^2 \right)} ,
\end{equation}
aggregate throughput
\begin{equation}
AT=\theta_1 + \theta_2 + \theta_3 ,
\end{equation}
and proportional fairness \cite{kelly1997crc}
\begin{equation}
PF=\log \theta_1 + \log \theta_2 + \log \theta_3.
\end{equation}
It can be observed that, in the steady state, scl-Aloha outperforms Aloha if $\epsilon$ in (\ref{eq:distributed}) is appropriately chosen.
We compare our protocol to optimally configured Aloha and not to CSMA, because CSMA is highly vulnerable to the hidden terminal problem that appears in multi-hop packet radio networks.
In our particular example, STA1 and STA3 cannot carrier-sense each other and they would continuously collide.

\begin{table*}[]
% increase table row spacing, adjust to taste
%\renewcommand{\arraystretch}{2.0}
%
\caption{Performance Comparison}
\label{tab:comparison}
\centering
\begin{tabular}{|c| |c|c|c|c|c|c|}
\hline
 Contention & \multicolumn{3}{|c|}{Per station throughput} & \multicolumn{3}{|c|}{Network performance}  \\
\hline
& $\theta_1$ & $\theta_2$ & $\theta_3$ & JF & AT & PF \\
\hline
\hline
 Aloha & 0.056 & 0.120 & 0.108 & 0.921 & 0.283 & -7.234 \\
 scl-Aloha & $\frac{1}{4(1+\epsilon)}$ & $\frac{1}{4(1+\epsilon)}$ & $\frac{1}{4(1+\epsilon)}$ & 1 & $\frac{3}{4(1+\epsilon)}$ & $-4.159 - 3\log(1+\epsilon)$ \\
\hline
\end{tabular}
\end{table*}

\begin{figure}
\centering
  \includegraphics[width=\linewidth]{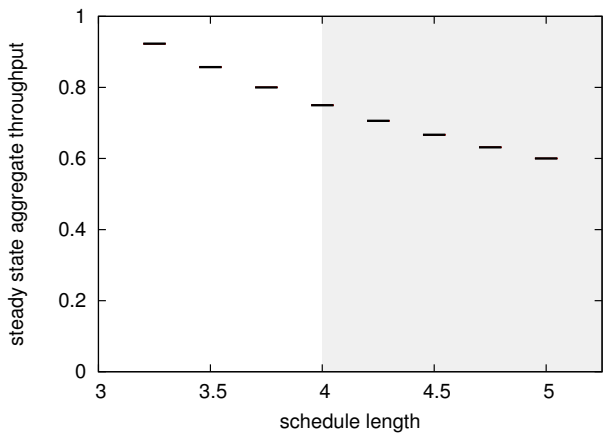}
\caption{Steady state aggregate traffic of the example network for different values of the schedule length.}
\label{fig:basic_plot}
\end{figure}

Fig.~\ref{fig:basic_plot} depicts the aggregate throughput of our example network for different schedule lengths ranging from 3.25 to 7 in increments of 0.25.
Throughput is measured in terms of successful transmissions per unit of time.
Although any schedule length which is strictly larger than three would be in principle feasible, the schedule length selection process described in the previous section restricts the possible values of the schedule length to those that are larger than four (four is the smaller power of two that is strictly larger than three).
The values larger than four are represented as a shaded area in the plot.
This shaded area covers all schedule lengths that are equal to \mbox{$4(1+\epsilon)$} for some positive value of epsilon.
The smaller the value of $\epsilon$, the larger the aggregate throughput.

\begin{figure}
\centering
  \includegraphics[width=\linewidth]{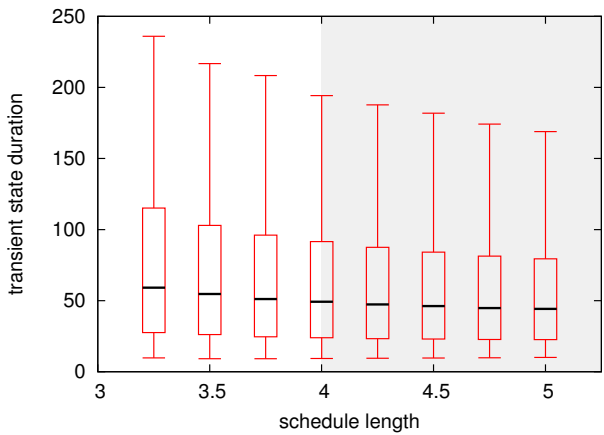}
\caption{A box-and-whiskers plot of the distribution of the absorption time for different schedule lengths. The 5, 25, 50, 75 and 95 percentiles are represented.}
\label{fig:transient_state_duration}
\end{figure}

In order to assess the duration of the transient state, we have used a simple simulator that builds upon the SimPy \cite{mueller2004doc} simulation framework for python
\footnote{This simulator includes all the assumptions and idealities discussed in the previous sections and it is available for download at www.jaumebarcelo.info/barcelo2011emp/simulator.py . Contact the first author for more details and the scripts required to generate the plot.}
.
We sweep schedule lengths from 3.25 to 5.00 in steps of 0.25.
In each step, we run one thousand simulations and compute the 5, 25, 50, 75, and 95 percentiles of the distribution of the duration of the transient state.
The results are plotted in Fig. \ref{fig:transient_state_duration}.
Again, the schedule length selection protocol restricts the possible values of the schedule to those that are larger than four, which correspond to the shaded area in the plot.

%It can be observed that the backoff protocol suggested in the previous section has a considerable shorter transient state that the one in \cite{barcelo2011cfo}.
%A possible explanation is that the aggressiveness of the protocol has been reduced.

\subsection{A six node ring topology}

\begin{figure}[]
\psfrag{d_1}[cc][cc]{$s_1$}
\psfrag{d_2}[cc][cc]{$s_2$}
\psfrag{d_3}[cc][cc]{$s_3$}
\psfrag{d_4}[cc][cc]{$s_4$}
\psfrag{d_5}[cc][cc]{$s_5$}
\psfrag{d_6}[cc][cc]{$s_6$}
\psfrag{f_1}[cc][cc]{$f_1$}
\psfrag{f_2}[cc][cc]{$f_2$}
\psfrag{f_3}[cc][cc]{$f_3$}
\psfrag{f_4}[cc][cc]{$f_4$}
\psfrag{f_5}[cc][cc]{$f_5$}
\psfrag{f_6}[cc][cc]{$f_6$}
\psfrag{l_1}[cc][cc]{$l_1$}
\psfrag{l_2}[cc][cc]{$l_2$}
\psfrag{l_3}[cc][cc]{$l_3$}
\psfrag{l_4}[cc][cc]{$l_4$}
\psfrag{l_5}[cc][cc]{$l_5$}
\psfrag{l_6}[cc][cc]{$l_6$}
\centering
\includegraphics[width=\linewidth]{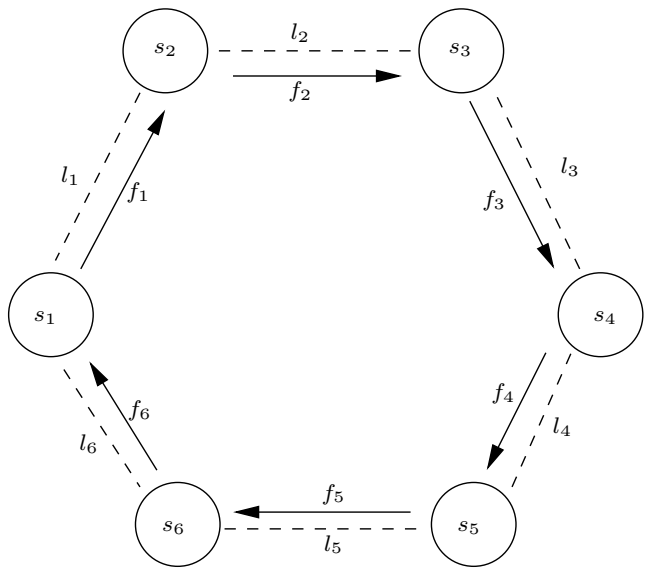}
\caption{A six node packet radio network.}
\label{fig:six_nodes}
\end{figure}

Fig. \ref{fig:six_nodes} represents a topology consisting of six nodes in a ring configuration.
In this configuration there is also a hidden node problem that cannot be prevented using CSMA.

\begin{figure}
\centering
  \includegraphics[width=\linewidth]{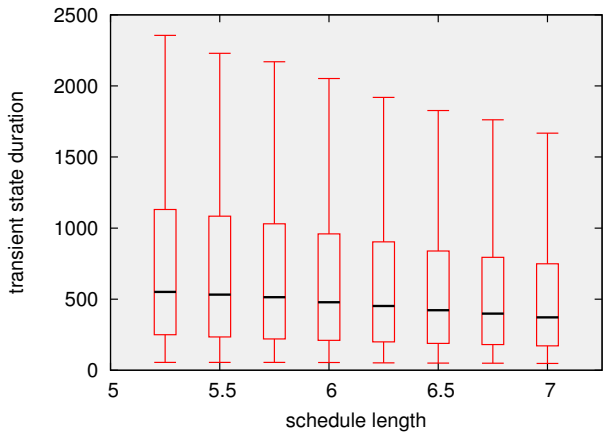}
\caption{A box-and-whiskers plot of the distribution of the absorption time for different schedule lengths. The 5, 25, 50, 75 and 95 percentiles are represented.}
\label{fig:six_nodes_transient_state_duration}
\end{figure}

We test this topology using simulation and derive statistics on the time required for the network to converge that are presented in Fig. \ref{fig:six_nodes_transient_state_duration}.
We consider schedule lengths ranging from 5.25 to 7 in increments of 0.25.

\begin{figure}
\centering
  \includegraphics[width=\linewidth]{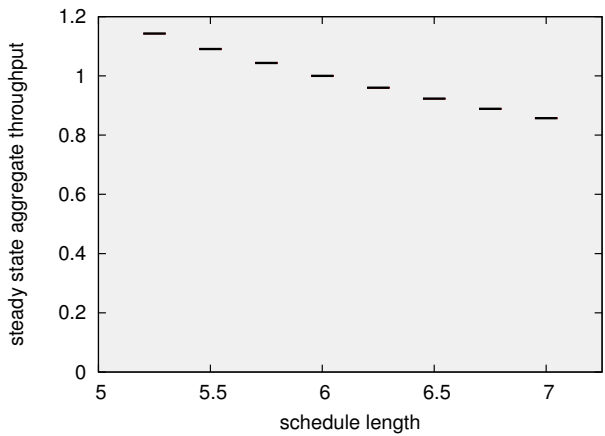}
\caption{Steady state aggregate traffic of the six node network for different values of the schedule length. Note that in this particular topology, it is possible that two stations simultaneously successfully transmit. For this reason, values of throughput higher than one are possible.}
\label{fig:throughput_in_steady_state}
\end{figure}

If optimal synchronization and scheduling were possible, two of the six stations could be transmitting at any time without interfering each other, which means an aggregate throughput equal to two.
The throughput in the steady state that is obtained by the proposed algorithm using the mentioned schedule lengths is roughly one (See Fig. \ref{fig:throughput_in_steady_state}).
Higher throughputs could be obtained at the expense of longer convergence times.

\begin{figure}
\centering
  \includegraphics[width=\linewidth]{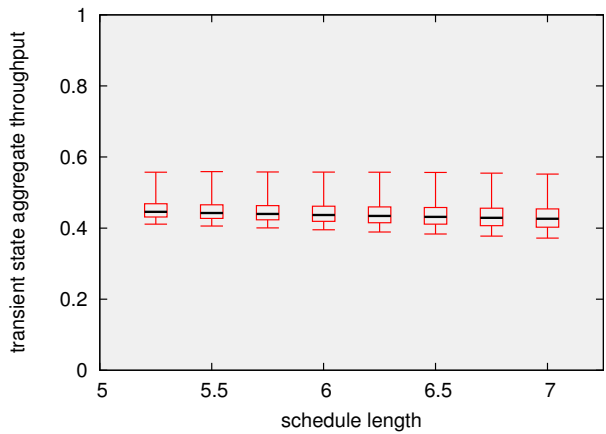}
\caption{Transient state aggregate traffic of the six node network for different values of the schedule length.}
\label{fig:throughput_in_transient_state}
\end{figure}

The performance that the network achieves during the transient state is substantially lower than the performance that is obtained in the steady state.
Fig. \ref{fig:throughput_in_transient_state} represents the statistics of the aggregate throughput that is achieved in the six node topology during the transient state.
Even though the presence of some collisions has a clear impact on performance, it does not present the starvation problems associated to hidden node scenarios.

\section{Refinements and future work}
\label{sec:refinements}

In this paper we have considered a very simple protocol that schedules the next transmission as a function of the result (either success or failure) of the last transmission.
We have seen that, despite its simplicity, this protocol can reach collision-free operation and solve the MAC layer scheduling problem in multi-hop packet radio networks.
A more sophisticated approach could converge more quickly to the collision-free schedule by including two additional rules in the protocol:
\begin{itemize}
\item Do not transmit while a packet is being received.
\item Do not transmit to a station that it is engaged in a transmission.
\end{itemize}
Violating any of those two rules results in a certain collision.

From this perspective, the results presented in this paper should be considered a lower bound on the performance that can be achieved by the family of collision-free MAC protocols.

As a first step towards collision-free protocols with better convergence properties, we have repeated the experiments with the six node ring topology using a hybrid protocol that combines both CSMA and the collision-free properties that have been described in the present paper.
In particular, if a node senses the medium busy after finishing a backoff, it starts a new backoff of exponentially distributed length.

\begin{figure}
\centering
  \includegraphics[width=\linewidth]{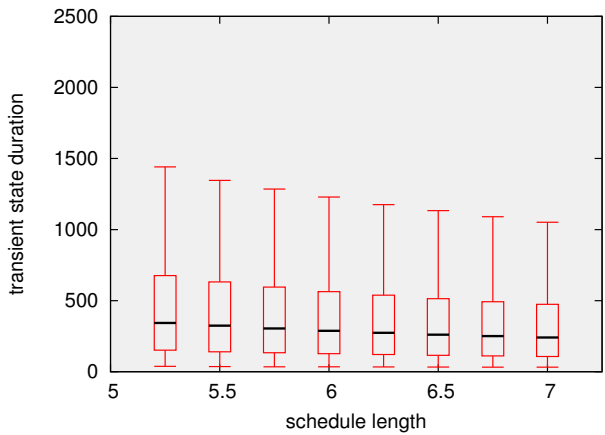}
\caption{Transient state duration statistics for the six node network and different values of the schedule length. The protocol combines CSMA and collision-free properties.}
\label{fig:transient_state_duration_hybrid}
\end{figure}

\begin{figure}
\centering
  \includegraphics[width=\linewidth]{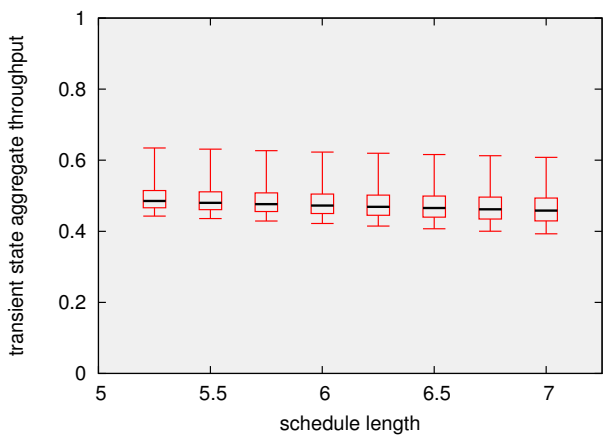}
\caption{Transient state aggregate traffic of the six node network for different values of the schedule length. The protocol combines CSMA and collision-free properties.}
\label{fig:throughput_in_transient_state_hybrid}
\end{figure}

The results are presented in Figs. \ref{fig:transient_state_duration_hybrid} and \ref{fig:throughput_in_transient_state_hybrid} and evidence a performance improvement, specially in terms of the time required to reach collision-free operation.
The throughput in the steady state is not plotted, since it is exactly the same as in Fig. \ref{fig:throughput_in_steady_state}.

A restriction of the protocol as presented in this paper is the fact that the number of flows in one station's neighborhood is assumed to be known, and therefore the nodes need to exchange \emph{hello} messages when the network starts to operate or when the traffic patterns change (as described in \cite{kar2004apf}).

Information about the number of flows in the neighborhood can be estimated by overhearing the neighbours' transmissions and ACKs.
The transmissions can be used to infer the number of outgoing flows while the ACKs make it possible to estimate the number of incoming flows.

In the presence of non-saturated flows, the stations with no packets to transmit might skip their turn while respecting the overall schedule.
If some of the flows have special requirements, such as requiring twice as much bandwidth as the normal flows, they should be split in two different flows (with the same origin and destination station as the original flow) and execute two different instances of the backoff protocol to have access to two transmission opportunities in each schedule.

A topology change or a change in the traffic pattern may require a re-computation of the schedule and the schedule length.
The applicability and the advantages of the protocol presented in this paper to dynamic networks is still a matter that requires further study.

Another aspect of interest is the combination of the proposed approach with existing mechanisms such as MCCA \cite{hiertz2010wms}.
Our protocol could be used to find a collision-free schedule for MCCA beacons, which in turn could be used to make channel reservations for data transmissions.

%The exploration of smarter learning protocols that converge quicker to collision-free operation, hybrid protocols that also consider reservation, and the implication of the use of estimated values for the number of incoming and outgoing flows in a neighborhood are interesting research topics for future work.

\section{Conclusion}
\label{sec:conclusion}

In this paper we have described the possibility of distributively constructing a collision-free schedule in multi-hop packet radio networks.
We have first discussed the problem of channel time fragmentation and explained that this problem can be alleviated by using delayed ACKs at the link layer.
Then we have dealt with the distributed computation of the schedule length, requiring only the exchange of local information among neighbours.
After the schedule length has been computed, the stations engage in the actual contention, using a deterministic backoff if a packet has been successfully transmitted (and acknowledged), and a random backoff otherwise.
When the collision-free schedule is reached, all the stations behave deterministically and no channel time is wasted in the form of collisions.
Therefore, the proposed protocol offers the possibility of making a more efficient use of the channel time.
The collision-free operation is reached only after a transient state, the duration of which we assessed for two particular topologies using simulation.

%Toroid. We use this surface to prevent border effects. The results that can be obtained are equivalent to the ones that would be obtained for the inner part of a large network.

\begin{acknowledgements}
The authors are thankful to the anonymous reviewers for their constructive comments that helped to substantially improve the final version of this manuscript.
This work has been partially funded by the Spanish Government (grant TEC2008-0655) and the European Commission (grant CIP-ICT PSP-2011-5).
The views expressed in this article are solely those of the authors and do not represent the views of the Spanish Government or the European Commission.
\end{acknowledgements}

% BibTeX users please use one of
%\bibliographystyle{spbasic}      % basic style, author-year citations
\bibliographystyle{spmpsci}      % mathematics and physical sciences
\bibliography{my_bib}   % name your BibTeX data base

\begin{thebibliography}{10}
\providecommand{\url}[1]{{#1}}
\providecommand{\urlprefix}{URL }
\expandafter\ifx\csname urlstyle\endcsname\relax
  \providecommand{\doi}[1]{DOI~\discretionary{}{}{}#1}\else
  \providecommand{\doi}{DOI~\discretionary{}{}{}\begingroup
  \urlstyle{rm}\Url}\fi

\bibitem{andreev2010ssv}
Andreev, K., Boyko, P.: {Simulation Study of VoIP Performance in IEEE 802.11
  Wireless Mesh Networks}.
\newblock MACOM pp. 139--150 (2010)

\bibitem{arikan1984scr}
Arikan, E.: {Some Complexity Results about Packet Radio Networks}.
\newblock IEEE Transactions on Information Theory \textbf{30}(4), 681--685
  (1984)

\bibitem{barcelo2008lba}
Barcelo, J., Bellalta, B., Cano, C., Oliver, M.: {Learning-BEB: Avoiding
  Collisions in WLAN}.
\newblock In: Eunice (2008)

\bibitem{barcelo2009cpa}
Barcelo, J., Bellalta, B., Cano, C., Sfairopoulou, A., Oliver, M.: {Carrier
  Sense Multiple Access with Enhanced Collision Avoidance: a Performance
  Analysis}.
\newblock In: ACM IWCMC (2009)

\bibitem{barcelo2011tcf}
Barcelo, J., Bellalta, B., Cano, C., Sfairopoulou, A., Oliver, M.: {Towards a
  Collision-Free WLAN: Dynamic Parameter Adjustment in CSMA/E2CA}.
\newblock In: Eurasip Journal on Wireless Communications and Networking (2011)

\bibitem{barcelo2009tpc}
Barcelo, J., Bellalta, B., Cano, C., Sfairopoulou, A., Oliver, M., Zuidweg, J.:
  {Traffic Prioritization for Carrience Sense Multiple Access with Enhanced
  Collision Avoidance}.
\newblock In: MACOM (2009)

\bibitem{barcelo2011cfo}
Barcelo, J., Bellalta, B., Oliver, M., Banchs, A.: {Collision Free Operation in
  Ad-Hoc Networks}.
\newblock In: MACOM (2011)

\bibitem{bellalta2009vtc}
Barcelo, J., Bellalta, B., Sfairopoulou, A., Cano, C., Oliver, M.: {CSMA with
  Enhanced Collision Avoidance: a Performance Assessment}.
\newblock In: IEEE VTC Spring (2009)

\bibitem{barcelo2010fcc}
Barcelo, J., Lopez-Toledo, A., Cano, C., Oliver, M.: {Fairness and Convergence
  of CSMA with Enhanced Collision Avoidance}.
\newblock In: ICC (2010)

\bibitem{chambers2002grr}
Chambers, B.: The grid roofnet: a rooftop ad hoc wireless network.
\newblock Ph.D. thesis, Massachusetts Institute of Technology (2002)

\bibitem{choi2005eei}
Choi, J., Yoo, J., Choi, S., Kim, C.: {EBA: An Enhancement of the IEEE 802. 11
  DCF via Distributed Reservation}.
\newblock IEEE Transactions on Mobile Computing \textbf{4}(4), 378--390 (2005)

\bibitem{cicconetti2008sdr}
Cicconetti, C., Lenzini, L., Mingozzi, E.: Scheduling and dynamic relocation
  for ieee 802.11 s mesh deterministic access.
\newblock In: IEEE SECON, pp. 19--27 (2008)

\bibitem{duffy2011dcs}
Duffy, K.R., Bordenave, C., Leith, D.J.: Decentralized constraint satisfaction.
\newblock CoRR \textbf{abs/1103.3240} (2011)

\bibitem{fang2010dlm}
Fang, M., Malone, D., Duffy, K., Leith, D.: {Decentralised Learning MACs for
  Collision-free Access in WLANs}.
\newblock Arxiv preprint arXiv:1009.4386v2  (2011)

\bibitem{gurewitz2009mmo}
Gurewitz, O., Mancuso, V., Shi, J., Knightly, E.: {Measurement and Modeling of
  the Origins of Starvation of Congestion-Controlled Flows in Wireless Mesh
  Networks}.
\newblock IEEE/ACM Transactions on Networking \textbf{17}(6), 1832--1845 (2009)

\bibitem{he2009sbr}
He, Y., Yuan, R., Sun, J., Gong, W.: {Semi-Random Backoff: Towards Resource
  Reservation for Channel Access in Wireless LANs}.
\newblock In: IEEE ICNP, pp. 21--30 (2009)

\bibitem{hiertz2010wms}
Hiertz, G., Denteneer, D., Max, S., Taori, R., Cardona, J., Berlemann, L.,
  Walke, B.: Ieee 802.11 s: The wlan mesh standard.
\newblock IEEE Wireless Communications Magazine \textbf{17}(1), 104--111 (2010)

\bibitem{hui2011epp}
Hui, K., Li, T., Guo, D., Berry, R.: Exploiting peer-to-peer state exchange for
  distributed medium access control.
\newblock In: IEEE ISIT, pp. 2368--2372. IEEE (2011)

\bibitem{jain1991acs}
Jain, R.: {The Art of Computer Systems Performance Analysis}.
\newblock John Wiley \& Sons New York (1991)

\bibitem{kar2004apf}
Kar, K., Sarkar, S., Tassiulas, L.: {Achieving Proportional Fairness Using
  Local Information in Aloha Networks}.
\newblock IEEE Transactions on Automatic Control \textbf{49}(10), 1858--1863
  (2004)

\bibitem{kelly1997crc}
Kelly, F.: Charging and rate control for elastic traffic.
\newblock European transactions on Telecommunications \textbf{8}(1), 33--37
  (1997)

\bibitem{khan2011sod}
Khan, Z., Lehtomaki, J., Mustonen, M., Matinmikko, M.: Sensing order dispersion
  for autonomous cognitive radios.
\newblock In: CROWNCOM, pp. 191--195. IEEE (2011)

\bibitem{krasilov2011iem}
Krasilov, A., Lyakhov, A., Safonov, A.: {Interference, Even with MCCA Channel
  Access Method in IEEE 802.11s Mesh Networks}.
\newblock In: MASS, pp. 752 --757 (2011)

\bibitem{martorell2012pec}
Martorell, G., Riera, F., Femenias, G., Barcelo, J., Bellalta, B.: {On the
  performance evaluation of CSMA/E2CA protocol with open loop ARF-based
  adaptive modulation and coding}.
\newblock In: European Wireless (2012)

\bibitem{mueller2004doc}
Mueller, K.: {SimPy Documentation}.
\newblock \url{http://simpy.sourceforge.net/} (2004).
\newblock [Online; accessed 05-December-2011]

\bibitem{oliver2010wca}
Oliver, M., Zuidweg, J., Batikas, M.: {Wireless Commons Against the Digital
  Divide}.
\newblock In: IEEE ISTAS (2010)

\bibitem{ramanathan1993sam}
Ramanathan, S., Lloyd, E.: {Scheduling Algorithms for Multihop Radio Networks}.
\newblock IEEE/ACM Transactions on Networking \textbf{1}(2), 166--177 (1993)

\bibitem{sommer2009gcs}
Sommer, P., Wattenhofer, R.: {Gradient Clock Synchronization in Wireless Sensor
  Networks}.
\newblock In: ACM/IEEE IPSN (2009)

\bibitem{tian2008ipc}
Tian, X., Chen, X., Ideguchi, T., Okuda, T.: {Improving protocol capacity by
  scheduling random access on WLANs}.
\newblock Telecommunication Systems \textbf{37}, 19--28 (2008)

\bibitem{tobagi1987mpa}
Tobagi, F.: {Modeling and Performance Analysis of Multihop Packet Radio
  Networks}.
\newblock Proceedings of the IEEE \textbf{75}(1), 135--155 (1987)

\bibitem{yi2010msl}
Yi, Y., De~Veciana, G., Shakkottai, S.: {MAC Scheduling With Low Overheads by
  Learning Neighborhood Contention Patterns}.
\newblock IEEE/ACM Transactions on Networking  (2010)

\end{thebibliography}

\end{document}